\documentclass[11pt,a4paper]{article}
  \usepackage{graphicx}
  \usepackage{amsmath, amssymb, graphics, setspace}
  \usepackage{calligra}
  \usepackage{soul}

  \usepackage[english]{babel}
  \usepackage[dvipsnames]{xcolor}

  \usepackage{geometry}
  \usepackage{marginnote}
  \usepackage{color}
  \usepackage{relsize}

  \newtheorem{theorem}{Theorem}[section]

  \providecommand{\keywords}[1]{\textbf{\textit{Keywords:\,\,}} #1}

\numberwithin{equation}{section}

\begin{document}
   
  \title{Dynamics of the $\Lambda$CDM model of the universe from     the aspect of the dynamical systems theory}
   \author{Danijela Brankovi\'c \\ 
   School of Electrical Engineering, University of Belgrade,\\
  Bulevar kralja Aleksandra 73, 11000 Belgrade, Serbia\\
  danijela@etf.bg.ac.rs\\ 
  \\
  \v Zarko Mijajlovi\'c \\
  Faculty of Mathematics, University of Belgrade, \\
  Studentski trg 16,  11000 Belgrade, Serbia \\
  zarkom@matf.bg.ac.rs\\}
  
  \date{}
  \maketitle 
  
  \begin{abstract}
  
  In this paper we exploit the theory of the dynamical systems to study the dynamics of the standard cosmological model of the universe, which is known as the $\Lambda$CDM model. We assume that the matter content in our universe consists of barotropic perfect fluids without mutual interaction. Furthermore, we present the appropriate physical interpretation, as well as new dependencies between the scale expansion factor of the universe and cosmological density parameters.
  \end{abstract}
  
  \noindent
% \PACS{PACS code1 \and PACS code2 \and more}
%\subclass{34A34 \and 37N99 \and 83F05}
  
 \section{Introduction}
\label{intro}

Our first goal is to prove that the system of the Friedmann equations is equivalent to the system of the first order nonlinear differential equations whose dependent variables are exactly density parameters of the material in the universe. Furthermore, we solve the obtained system in order to get a new parametrization of density parameters with respect to the scale expansion factor of the universe.
  
  Our starting point is the system of the Friedmann equations \cite{friedmann} with the cosmological constant $\Lambda$
  \begin{equation}\label{sf}
  	\begin{array}{l}
  		\smallskip
  		 \displaystyle \left(\dfrac{\dot{a}}{a}\right)^2=\dfrac{8\pi G}{3}\rho - 		 					  \dfrac{kc^2}{a^2}+\frac{\Lambda c^2}{3},\\	
  		
  	    \smallskip
  		\displaystyle \dfrac{\ddot{a}}{a}=-\dfrac{4\pi G}{3}\left(\rho+\dfrac{3p}							 {c^2}\right)+\dfrac{\Lambda c^2}{3},\\
  		
  		\smallskip 		
  		\displaystyle \dot{\rho}+3\dfrac{\dot a}{a}\left(\rho+\frac{p}{c^2}							      \right)=0,  	
  	\end{array}
  \end{equation}
 consisting of nonlinear differential equations of the first and second order that describes the dynamics of our universe. The first equation in (\ref{sf}) is called the Friedmann equation, while the second and the third one are refered to the acceleration equation and the fluid equation (also known as conservation equation), respectively. Cosmological parameters $a$, $\rho$ and $p$ that appear in (\ref{sf}) are functions of time variable $t$ and they represent the scale expansion factor of the universe, the average density of matter content in the universe and the pressure of matter content in the universe, respectively. The curvature of the universe is denoted with $k$. In this paper we discuss only flat universe ($k=0$) and open universe ($k < 0$). Derivatives with respect to $t$ are denoted with a dot. Considering physical limitations, we assume that all functions that appear in this paper are continuously differentiable as many times as needed.
  
  Only two equations from (\ref{sf}) are independent, since the Friedmann equation and the acceleration equation imply the fluid equation, as well as the acceleration equation directly follows from the Friedmann equation and the fluid equation. Therefore the system (\ref{sf}) consists from two independent equations that contain three functions $a(t)$, $\rho (t)$ and $p(t)$ that are unknown. In order to have a determined system, standard approach is assuming an equation of state for the matter content that builds up the universe. Here we consider that the matter components in our universe are barotropic perfect fluids ("matter" and "radiation") which do not interact with each other, more precisely that a single fluid component (for example "matter") do not interact with the other fluid component ("radiation"). By "matter" it is typically considered baryonic matter plus cold dark matter, while "radiation" refers to relativistic particles, such as neutrinos and photons.
  
The equation of state for barotropic perfect $i$-fluid is a linear relation between $p_i$ and $\rho_i$
  \begin{equation}\label{EoS}
  	\displaystyle p_i = \omega_i \rho_i c^2,
  \end{equation}
where $\rho_i$ is the average density, $p_i$ is the pressure and $\omega_i$ is the equation of state parameter of $i$-fluid \cite{ds}, \cite{relativity}. In this paper we consider that $\omega_i$ is a constant from the interval $\left[ 0,1\right]$, since that is consistent with the macroscopic physics \cite{ds}.

 The assumption that a single fluid component do not interact with the other fluid component is reflecting in that the fluid equation is conserved for every $i$-fluid \cite{ds}, \cite{jungle}, i.e. 
	\begin{equation}\label{ff}		
  		\displaystyle \dot{\rho_i}+3\dfrac{\dot a}{a}\left(\rho_i+\dfrac{p_i}{c^2}						  \right)=0.
	\end{equation}
Substituting (\ref{EoS}) and the Hubble parameter $H(t)={\dot a (t)}/{a(t)}$ in (\ref{ff}), we infer the following form of the fluid equation for $i$-fluid
	\begin{equation}\label{fluidf}
  		\displaystyle \dot{\rho_i}=-3H\rho_i\left(1+\omega_i\right).
	\end{equation}
The equation (\ref{fluidf}) in the terms of $\rho_m$ (the average density of $m$-fluid (matter)) and $\rho_r$ (the average density of $r$-fluid (radiation)), is in the form
	\begin{equation}\label{mr}
		\begin{array}{l}		
  			\displaystyle \dot{\rho_m}=-3H\rho_m\left(1+\omega_m\right),\\
  			\displaystyle \dot{\rho_r}=-3H\rho_r\left(1+\omega_r\right).
  		\end{array}
	\end{equation}
	
Clearly, $\rho_m$ and $\rho_r$ contribute to the average density of the matter content in the universe, as well as $p_m$ and $p_r$ contribute to the pressure of the matter content in the universe, with 
	\begin{equation}\label{dp}
		\begin{array}{l}		
  			\displaystyle \rho=\rho_m+\rho_r,\\
  			\displaystyle p=p_m+p_r.
  		\end{array}
	\end{equation}
	
Cosmological parameters that play a central role in this manuscript are density parameters of the material in the universe, defined as (see \cite{liddle})
	\begin{equation}\label{Om}
		\begin{array}{l}
			\smallskip
			\displaystyle \Omega_m(t)=\dfrac{8\pi G}{3 H^2(t)}\rho_m(t),\\
			
			\smallskip
			\displaystyle \Omega_k(t) =-\dfrac{k c^2}{a^2(t)H^2(t)},\\
			
			\smallskip
			\displaystyle \Omega_r(t)=\dfrac{8\pi G}{3 H^2(t)}\rho_r(t),\\
			
			\smallskip
			\displaystyle \Omega_{\Lambda}(t)=\dfrac{\Lambda c^2}{3H^2(t)},
		\end{array}
	\end{equation}
where $\Omega_m(t)$, $\Omega_k(t)$, $\Omega_r(t)$ and $\Omega_{\Lambda}(t)$ denote matter density, spatial curvature density, radiation density and cosmological constant density. Throughout the paper, $\Omega_i$ will stand for any of the following density parameters $\Omega_m(t)$, $\Omega_k(t)$, $\Omega_r(t)$, $\Omega_{\Lambda}(t)$, unless otherwise noted.

Acording to the $\Lambda$CDM model, the universe's evolution consists of three main epochs. The behaviour of the universe in each epoch is governed by one dominant component. Radiation domination was the first epoch, that lasted until around $50 \ 000$ yrs. For the next $350 \ 000$ yrs, radiation and matter almost equally occupied the universe. After recombination era (around $400 \ 000$ yrs), matter was the dominant fluid in the universe. Afterwards, at approximately $10^9$ yrs, matter and dark energy density were almost equal. Finally, dark energy that is represented by the cosmological constant, has become the dominant component in the universe, since approximately $7.7 \cdot 10^9$ years after the Big Bang.
 
\section{Friedmann equations as a dynamical system}
\label{sec:derivation}
  
By dividing the Friedmann equation in (\ref{sf}) with $H^2 \neq 0$ ($H=0$ implies static universe) and substituting $\rho=\rho_m+\rho_r$, we obtain
	\begin{equation}\label{sum}
		\Omega_m+\Omega_r+\Omega_k+\Omega_{\Lambda}=1.
	\end{equation}

According to (\ref{Om}) and (\ref{sum}), as well as the assumption that $k\leq 0$, it is clear that 
	\begin{equation}
		 \ 0 \leq \Omega_i \leq 1.
	\end{equation} 

From (\ref{sum}) directly follows
	\begin{equation}\label{Omk}
		\displaystyle \Omega_k=1-\Omega_m-\Omega_r-\Omega_{\Lambda}.
	\end{equation}	
	
In order to infer the dynamical system, we present a method which is pictured in various literature (for example see \cite{ds}, \cite{jungle}, \cite{garcia}, \cite{uzan}, \cite{goliath}) in a somewhat different way.

We perform change of variables $\xi(t)=\ln(a(t))$, denoting the derivatives with respect to $\xi$ with $'$. Note that $\xi$ better approximates linearity of time $t$ than $a(t)$. Differentiating the density parameters $\Omega_m$, $\Omega_r$ and $\Omega_{\Lambda}$ with respect to $\xi$, we infer
	\begin{equation}\label{prim}
		\displaystyle {\Omega_i}'=\dfrac{d \Omega_i}{dt} \cdot \dfrac{dt}{d\xi}= 				\dfrac{\dot{\Omega_i}}{H}.
	\end{equation}
	
Substituting $\displaystyle \Omega_i=8\pi G \rho_i/{3 H^2}$  and (\ref{fluidf}) in (\ref{prim}), for $i \in \left\{m, r \right\}$ we obtain	
	\begin{equation}\label{oi}
		\displaystyle {\Omega_i}'=-3\Omega_i (1+\omega_i)+\Omega_i \cdot 						\frac{-2\dot{H}}{H^2}.
	\end{equation}
	
If we replace $\displaystyle \rho=\rho_m+\rho_r$ and $H={\dot a}/{a}$ in the Friedmann equation in (\ref{sf}) and then differentiate it with respect to the time $t$, we infer
	\begin{equation}\label{step}
		\displaystyle 2H\dot{H}=	\dfrac{8\pi G}{3} \left(\dot{\rho_m} +\dot{\rho_r}			\right)+2kc^2 \cdot \dfrac{\dot{a}}{a^3}.	
	\end{equation}
	
Furthermore, substitution of (\ref{mr}), (\ref{Om}) and (\ref{Omk}) in (\ref{step}) transforms previous equation into
	\begin{equation}\label{Hpart}
		\displaystyle \dfrac{-2\dot{H}}{H^2}=2+\Omega_m \left(1+3\omega_m \right)+			\Omega_r \left(1+3\omega_r \right)-2\Omega_{\Lambda}.
	\end{equation}
	
Replacing (\ref{Hpart}) into (\ref{oi}), for $i \in \left\{m, r \right\}$ we get
	\begin{equation}\label{dsmr}
		\displaystyle {\Omega_i}'=-3\Omega_i (1+\omega_i)+\Omega_i \left(2+\Omega_m 			\left(1+3\omega_m \right)+\Omega_r \left(1+3\omega_r 									\right)-2\Omega_{\Lambda}\right).		
	\end{equation}
	
For $\Omega_{\Lambda}=\Lambda c^2/{3H^2}$, we have
	\begin{equation}
		\displaystyle {\Omega_{\Lambda}}'=\Omega_{\Lambda}\cdot \frac{-2\dot{H}}{H^2}, 
	\end{equation}	 	
wherefrom, using (\ref{Hpart}), follows
	\begin{equation}\label{dsL}
		\displaystyle {\Omega_{\Lambda}}'=\Omega_{\Lambda} \left(2+\Omega_m 					\left(1+3\omega_m \right)+\Omega_r \left(1+3\omega_r 									\right)-2\Omega_{\Lambda}\right).	
	\end{equation}	 

Gathering (\ref{dsmr}) and (\ref{dsL}) together, we finally infer the dynamical system
	\begin{equation}\label{ds}
		\begin{array}{l}
			\smallskip
			\displaystyle {\Omega_m}'=\Omega_m \left(-\left( 1+3\omega_m \right)+											 \Omega_m \left(1+3\omega_m \right)+\Omega_r 											 \left(1+3\omega_r \right)-2\Omega_{\Lambda} 											 \right),\\	
			
			\smallskip
			\displaystyle {\Omega_r}'=\Omega_r \left(-\left( 1+3\omega_r \right)+											 \Omega_m \left(1+3\omega_m \right)+\Omega_r 											 \left(1+3\omega_r \right)-2\Omega_{\Lambda}											 \right),\\
			
			\smallskip
			\displaystyle {\Omega_{\Lambda}}'=\Omega_{\Lambda} \left(2+\Omega_m 													 \left(1+3\omega_m \right)+\Omega_r 													 \left(1+3\omega_r 																	 \right)-2\Omega_{\Lambda}\right),
		\end{array}
	\end{equation}
which is the system of the first order nonlinear differential equations whose dependent variables are density parameters of the material in the universe.

In Theorem~\ref{theorem 1} we prove that the system of the Friedmann equations (\ref{sf}) is equivalent to the system of the first order nonlinear differential equations (\ref{ds}) under certain conditions. 

\begin{theorem}\label{theorem 1}
		Assuming $\displaystyle p_m = \omega_m \rho_m c^2$, $\displaystyle p_r = \omega_r \rho_r c^2$  and the identities (\ref{dp}), the system of the Friedmann equations (\ref{sf}) together with the equations (\ref{mr}) are equivalent to the system 			
		\begin{equation}\label{system}
		\begin{array}{l}
			\smallskip
			\displaystyle {\Omega_m}'=\Omega_m \left(-\left( 1+3\omega_m \right)+											 \Omega_m \left(1+3\omega_m \right)+\Omega_r 											 \left(1+3\omega_r \right)-2\Omega_{\Lambda} 											 \right),\\	
			
			\smallskip
			\displaystyle {\Omega_r}'=\Omega_r \left(-\left( 1+3\omega_r \right)+											 \Omega_m \left(1+3\omega_m \right)+\Omega_r 											 \left(1+3\omega_r \right)-2\Omega_{\Lambda}											 \right),\\
			
			\smallskip
			\displaystyle {\Omega_{\Lambda}}'=\Omega_{\Lambda} \left(2+\Omega_m 													 \left(1+3\omega_m \right)+\Omega_r 													 \left(1+3\omega_r 																	 \right)-2\Omega_{\Lambda}\right),\\
			
			\smallskip
			\displaystyle \Omega_m+\Omega_r+\Omega_{\Lambda}+\Omega_k=1,\\
			
			\smallskip
			{\Omega_i}'=\dfrac{\dot{\Omega_i}}{H},				
		\end{array}
	\end{equation}
where $\Omega_i$ are defined in (\ref{Om}).
	\end{theorem}
  
\noindent{\bf{Proof}} 
We already proved that from the system (\ref{sf}) and the equations (\ref{mr}) under the assumptions from Theorem~\ref{theorem 1} follows the system (\ref{system}). Now we prove the opposite direction.

From $\displaystyle \Omega_m+\Omega_r+\Omega_{\Lambda}+\Omega_k=1$, definitions of the density parameters in (\ref{Om}) and the first identity in (\ref{dp}) directly follows the Friedmann equation in (\ref{sf}). The next step is authenticating the equations (\ref{mr}). 

We turn our attention to the system (\ref{system}).

From $\displaystyle {\Omega_{\Lambda}}'=\dot{\Omega_{\Lambda}}/H$ and the third equation in (\ref{system}) follows
	\begin{equation}\label{sec}
		\displaystyle \dfrac{-2\dot{H}}{H^2}=2+\Omega_m \left(1+3\omega_m \right)+			\Omega_r \left(1+3\omega_r \right)-2\Omega_{\Lambda}.
	\end{equation}

Since $\displaystyle {\Omega_i}'=\dot{\Omega_i}/H$, for $i \in \left\{m, r \right\}$ we have
	\begin{equation}
		\displaystyle {\Omega_i}'=\dfrac{8\pi G}{3H^3} \dot{\rho_i}+\dfrac{8\pi G 			\rho_i}{3H^2} \cdot \dfrac{-2\dot{H}}{H^2},
	\end{equation}
wherefrom, substituting (\ref{sec}) and $\displaystyle \Omega_i=8\pi G \rho_i/{3 H^2}$, we infer
	\begin{equation}
		\displaystyle \dfrac{{\Omega_i}'}{\Omega_i}=\dfrac{\dot{\rho_i}}{\rho_i}				\cdot \dfrac{1}{H}+2+\Omega_m \left(1+3\omega_m \right)+ \Omega_r 					\left(1+3\omega_r \right)-2\Omega_{\Lambda}.
	\end{equation}
	
If we compare previous equation for $i \in \left\{m, r \right\}$ with the first two equations in the system (\ref{system}), we conclude
	\begin{equation}\label{st1}
		\begin{array}{l}
			\smallskip
			\displaystyle \dfrac{\dot{\rho_m}}{\rho_m}	\cdot \dfrac{1}{H}+2=-					(1+3\omega_m),\\
			
			\smallskip
			\displaystyle \dfrac{\dot{\rho_r}}{\rho_r}	\cdot \dfrac{1}{H}+2=-					(1+3\omega_r).			
		\end{array}								
	\end{equation}

From (\ref{st1}) easy follows (\ref{mr}). Using the assumptions $\displaystyle p_m = \omega_m \rho_m c^2$, $\displaystyle p_r = \omega_r \rho_r c^2$, the identities (\ref{dp}) as well as (\ref{mr}), fluid equation in (\ref{sf}) is immediately obtained. Since we already mentioned that from the Friedmann equation and fluid equation follows acceleration equation, we obtained the equivalence with the system (\ref{sf}).
\hfill $\Box$
\medskip

The obtained equivalence guarantees that dynamics of the systems (\ref{system}) and (\ref{sf}) are the same, under standard assumptions in cosmology. We note that in literature the same conclusion is used but only with $\omega_m=0$ and $\omega_r=1/3$, however only one implication is inferred, not the equivalence.

We consider some natural questions that arise here. For example, can we obtain analytical solution of the system (\ref{system})? Under what circumstances is that possible? What information will that solution provide us with? The answer to these questions we present in the Sect.~\ref{sec:solutions}.

\section{Solutions of the dynamical system} 
\label{sec:solutions}

In Sect.~\ref{sec:solutions} we solve the system (\ref{system}) in various cases, depending on whether some of the density parameters $\Omega_i$ are equal to zero. Every case that we analyze here corresponds to some model of the universe. More precisely, it represents some stage, or perhaps some medium stage in the evolution of the universe. To each case we give the appropriate physical interpretation \cite{cosmictime}. 

\subsection{Three dimensional dynamical system}
\label{sec:3}

Here we consider that $\Omega_m>0$, $\Omega_r>0$, $\Omega_{\Lambda}>0$ and $\Omega_k>0$. Therefore, this case represents an open universe in the matter dominated epoch. Moreover, it represents an open universe in the phase of the transition between radiation and dark energy that occurred approximately around $5 \cdot 10^8$ yrs after the Big Bang \cite{dz}. 

First, we rewrite the system (\ref{system}) in the following form
	\begin{equation}\label{systemG}
		\begin{array}{l}
			\smallskip
			\displaystyle {\Omega_m}'=\Omega_m \left(-(1+3\omega_m)+(1+3\omega_r)					\Omega_r -2\Omega_{\Lambda}\right)+(1+3\omega_m){\Omega_m}^2,\\	
			
			\smallskip
			\displaystyle {\Omega_r}'=\Omega_r \left(-(1+3\omega_r)+(1+3\omega_m)					\Omega_m -2\Omega_{\Lambda}\right)+(1+3\omega_r){\Omega_r}^2,\\
			
			\smallskip
			\displaystyle {\Omega_{\Lambda}}'=\Omega_{\Lambda} \left(2+(1+3\omega_m)				\Omega_m+(1+3\omega_r)\Omega_r\right)-2{\Omega_{\Lambda}}^2,\\		
		\end{array}				
	\end{equation}		  
with $\displaystyle \Omega_m+\Omega_r+\Omega_{\Lambda}+\Omega_k=1$ and 
${\Omega_i}'=\dfrac{\dot{\Omega_i}}{H}$. Since $\Omega_m$, $\Omega_r$, $\Omega_{\Lambda}$ and $\Omega_k$ are strictly positive and their sum is equal to 1, hence all density parameters are less than 1.

Note that the density parameters can be observed as a functions of $\xi=\ln a$. 
To simplify the system (\ref{systemG}), we introduce new variables with
	\begin{equation}\label{changeG}
		z_m(\xi)=1/{\Omega_m (\xi)}, \quad z_r(\xi)=1/{\Omega_r(\xi)}, \quad 			z_{\Lambda}(\xi)=1/{\Omega_{\Lambda}(\xi)}.
	\end{equation}

Replacing the density parameters in (\ref{systemG}) with new variables $z_m$, $z_r$ and $z_{\Lambda}$, we infer less complicated system of equations
	\begin{equation}\label{systemGz}
		\begin{array}{l}
			\smallskip
			\displaystyle {z_m}'=-z_m \left(-(1+3\omega_m)+\dfrac{1+3\omega_r}					{z_r}-\dfrac{2}{z_{\Lambda}}\right)-(1+3\omega_m),\\	
			
			\smallskip
			\displaystyle {z_r}'=-z_r \left(-(1+3\omega_r)+\dfrac{1+3\omega_m}					{z_m}-\dfrac{2}{z_{\Lambda}}\right)-(1+3\omega_r),\\
			
			\smallskip
			\displaystyle {z_{\Lambda}}'=-z_{\Lambda} \left(2+\dfrac{1+3\omega_m}				{z_m}+\dfrac{1+3\omega_r}{z_r}\right)+2.\\		
		\end{array}				
	\end{equation}		  
The next step is solving the system (\ref{systemGz}) by the elimination method. 

The second and the third derivative of $z_m$ with respect to the variable $\xi$, after substituting the expressions for ${z_m}'$, ${z_r}'$ and ${z_{\Lambda}}'$ from (\ref{systemGz}), are in the form
	\begin{equation}\label{Gst}
		\begin{array}{l}
			\smallskip
			\displaystyle {z_m}''=-(1 + 3 \omega_m)^2  
 			-\dfrac{z_m \left( 1+6\omega_m-3\omega_r \right)\left( 1 + 							3\omega_r \right)}{z_r} \\
 			\hspace{12mm} + \dfrac{z_m\left( 8+12\omega_m +\left( 1+3\omega_m 					\right)^2 z_{\Lambda}\right)}{z_{\Lambda}},\\
			
			\smallskip
			\displaystyle {z_m}'''=-\left( 1+3\omega_m \right)^3 \\
			\hspace{12mm} -\dfrac{z_m \left( 1+3\omega_r \right) \left( 							1+27{\omega_m}^2+\omega_m \left(9-27\omega_r \right) 									-3\omega_r+9{\omega_r}^2 \right)}{z_r}\\
			 \hspace{12mm} +\dfrac{z_m\left( 26+72\omega_m+54{\omega_m}^2+\left( 1 				+3\omega_m\right)^3 z_{\Lambda}\right)}{z_{\Lambda}}.\\		
		\end{array}			
	\end{equation}
	
Now, we want to express $z_r$ and $z_{\Lambda}$ in the terms of $z_m$, ${z_m}'$ and  ${z_m}''$. In order to do so, we observe the function $F=F\left(z_m,{z_m}',{z_m}'',z_r,z_{\Lambda} \right)$, given with 
	\begin{equation}\label{bigF}
		\displaystyle F\left(z_m,{z_m}',{z_m}'',z_r,z_{\Lambda} \right)=\left( F_1 \left( z_m,{z_m}',z_r,z_{\Lambda} \right), F_2\left( z_m,{z_m}'',z_r,z_{\Lambda} \right) \right),
	\end{equation}
where
	\begin{equation}\begin{array}{l}
		\smallskip
		\displaystyle F_1={z_m}'+z_m \left(-(1+3\omega_m)+\dfrac{1+3\omega_r}					{z_r}-\dfrac{2}{z_{\Lambda}}\right)+(1+3\omega_m)=0,\\	
			
		\smallskip	
		\displaystyle F_2={z_m}''+(1 + 3 \omega_m)^2  
 			+\dfrac{z_m \left( 1+6\omega_m-3\omega_r \right)\left( 1 + 							3\omega_r \right)}{z_r} \\
 			\hspace{12mm} - \dfrac{z_m\left( 8+12\omega_m +\left( 1+3\omega_m 					\right)^2 z_{\Lambda}\right)}{z_{\Lambda}}=0.\\
	\end{array}
	\end{equation}
According to $z_m = 1/{\Omega_m} >1$, $\omega_r \in \left[ 0,1 \right]$ and our assumption that all functions that appear in this paper are continuously differentiable as many times as needed, as well as
	\begin{equation}
		\displaystyle \left|\begin{matrix}
		\dfrac{\partial F_1}{\partial z_r} & \dfrac{\partial F_1}{\partial z_{\Lambda}}\\
		\dfrac{\partial F_2}{\partial z_r} & \dfrac{\partial F_2}{\partial z_{\Lambda}}\\  
			\end{matrix}\right|=-\dfrac{6\left(1+3\omega_r\right)\left(1+\omega_r\right){z_m}^2}{{z_r}^2{z_{\Lambda}}^2} \neq 0,	
	\end{equation}
the implicit function theorem allows us to $z_r$ and $z_{\Lambda}$ express as functions of $z_m$, ${z_m}'$ and  ${z_m}''$, i.e. to infer the following	
	\begin{equation}\label{Gzrl}
		\begin{array}{l}
			\smallskip
			\displaystyle z_r=\dfrac{3z_m\left( 1+4\omega_r+3{\omega_r}^2 \right)}				{3\left( 1+4\omega_m+3{\omega_m}^2 \right)\left( z_m-1 \right)
			 -{z_m}' \left( 4+ 
			6\omega_m \right)+{z_m}''},\\
			 			
			\smallskip
			\displaystyle z_{\Lambda}=-\dfrac{6 z_m\left( 1+\omega_r \right)}{
			3\left(1+3\omega_m \right)\left( \omega_m-\omega_r \right)\left( 1-z_m 				\right)+{z_m}'\left( 1+6\omega_m -3\omega_r \right)-{z_m}''},\\			
		\end{array}
	\end{equation}	  
if the denominators of $z_r$ and $z_{\Lambda}$ are not equal to zero, i.e. if  

\begin{equation}\label{denominators}
			\displaystyle z_m  \neq 1+c_1 \mathrm{e}^{3 \left( 1+\omega_m \right) \xi} + c_2 \mathrm{e}^{\left( 1+3\omega_m \right) \xi} \ \ \mathrm{and} \ \
			\displaystyle z_m  \neq 1+c_3 \mathrm{e}^{3 \left(\omega_m - \omega_r \right) \xi} + c_4 \mathrm{e}^{\left( 1+3\omega_m \right) \xi},
	\end{equation}	
where $c_1$, $c_2$, $c_3$ and $c_4$ are arbitrary constants. Since $z_m \neq 0$ and $\omega_r \in \left[ 0,1 \right]$, from (\ref{Gzrl}) follows $z_r \neq 0$ and $z_{\Lambda} \neq 0$. If at least one of the conditions from (\ref{denominators}) is not satisfied, from (\ref{Gzrl}) we infer $\Omega_r=1/{z_r}=0$ or $\Omega_{\Lambda}=1/{z_{\Lambda}}=0$, which is not consistent with our assumptions in this case. The solutions for the cases  when only one of the density parameters is equal to zero are analyzed in the Subsect.~\ref{sec:2}. Concretely, the solution for $\Omega_r=0$ and the solution for $\Omega_{\Lambda}=0$ are obtained in the equations (\ref{sol22}) and (\ref{sol23}), respectively.
	
Substituting $z_r$ and $z_{\Lambda}$ from (\ref{Gzrl}) into ${z_m}'''$ from (\ref{Gst}) gives us following nonhomogeneous linear differential equation of the third order with constant coefficients 
	\begin{equation*}
		\begin{array}{l}
		\smallskip
		\displaystyle {z_m}'''-\left( 4+9\omega_m-3\omega_r \right){z_m}''+ 3\left( 			1+8\omega_m+9{\omega_m}^2 -4\omega_r-6\omega_m \omega_r \right){z_m}' \\
		
		\smallskip
		-9\left( 1+4\omega_m+3{\omega_m}^2 \right)\left( \omega_m-\omega_r 					\right)z_m+9\left( 1+4\omega_m+3{\omega_m}^2 \right)\left( \omega_m-\omega_r 		\right)=0,		
			\end{array}
	\end{equation*}
whose general solution is
	\begin{equation}\label{Ggs}
		\displaystyle z_m(\xi)=1+ c_1 \mathrm{e}^{3\left( \omega_m-\omega_r \right)\xi} 		+ c_2\mathrm{e}^{3\left( 1+\omega_m \right)\xi}
		+c_3\mathrm{e}^{\left( 1+3\omega_m \right)\xi}, 
	\end{equation}	
where $c_1$, $c_2$ and $c_3$ are strictly positive constants, so that $z_m=1/{\Omega_m}>1$.

Changing ${z_m}'(\xi)$ and ${z_m}''(\xi)$ into (\ref{Gzrl}) gives us $z_r(\xi)$ and $z_{\Lambda}(\xi)$. Collecting all together, solution of the system (\ref{systemGz}) is
	\begin{equation}\label{solGz}
		\begin{array}{l}
			\smallskip
			\displaystyle z_m(\xi)=1+c_1 \mathrm{e}^{3\left( \omega_m-\omega_r \right)				\xi}+c_2\mathrm{e}^{3\left( 1+\omega_m \right)\xi}
			+c_3\mathrm{e}^{\left( 1+3\omega_m \right)\xi}, \\
			
			\smallskip
			\displaystyle z_r(\xi)=\dfrac{\mathrm{e}^{3\left( \omega_r -\omega_m				\right)\xi}+c_1+c_2\mathrm{e}^{3\left( 1+\omega_r \right)\xi}
			+c_3\mathrm{e}^{\left( 1+3\omega_r \right)\xi}}{c_1},\\
			
			\smallskip
			\displaystyle z_{\Lambda}(\xi)=\dfrac{\mathrm{e}^{-3\left( 1+\omega_m 				\right)\xi}+c_1\mathrm{e}^{-3\left( 1+\omega_r \right)\xi}+c_2
			+c_3\mathrm{e}^{-2\xi}}{c_2},\\
		\end{array}
	\end{equation}
wherefrom, substituting (\ref{changeG}), we infer solutions of the system (\ref{systemG})
	\begin{equation}\label{solGOm}
		\begin{array}{l}
			\smallskip
			\displaystyle \Omega_m(\xi)=\dfrac{1}{1+c_1 \mathrm{e}^{3\left( \omega_m-				\omega_r \right)	\xi}+c_2\mathrm{e}^{3\left( 1+\omega_m \right)\xi}
			+c_3\mathrm{e}^{\left( 1+3\omega_m \right)\xi}}, \\
			
			\smallskip
			\displaystyle \Omega_r(\xi)=\dfrac{c_1}{\mathrm{e}^{3\left(				\omega_r  -\omega_m \right)\xi}+c_1+c_2\mathrm{e}^{3\left( 1+\omega_r \right)\xi}
			+c_3\mathrm{e}^{\left( 1+3\omega_r \right)\xi}},\\
			
			\smallskip
			\displaystyle \Omega_{\Lambda}(\xi)=\dfrac{c_2}{\mathrm{e}^{-3\left( 1+				\omega_m \right)\xi}+c_1\mathrm{e}^{-3\left( 1+\omega_r \right)\xi}+c_2
			+c_3\mathrm{e}^{-2\xi}}.\\		
		\end{array}
	\end{equation}
It is easy to see from (\ref{solGOm}) that $\Omega_i \in \left(0,1\right)$.
From (\ref{solGOm}) and (\ref{Omk}) we infer the following dependences
	\begin{equation}\label{shortGxi}
		\begin{array}{l}
			\smallskip \displaystyle \Omega_r(\xi)=c_1 \mathrm{e}^{3\left( \omega_m-\omega_r 				\right)\xi}\Omega_m(\xi),\\
			
			\smallskip \displaystyle \Omega_{\Lambda}(\xi)=c_2 \mathrm{e}^{3 \left( 1+\omega_m 				\right)\xi}\Omega_m(\xi),\\
			
			\smallskip \displaystyle \Omega_k(\xi)=1-\Omega_m(\xi) \left( 							1+c_1\mathrm{e}^{3\left( \omega_m-\omega_r \right)\xi}+c_2 \mathrm{e}^{3 \left( 1+\omega_m 				\right)\xi} \right).									
		\end{array}		
	\end{equation}
Changing $\xi=\ln a$ in (\ref{solGOm}) and (\ref{shortGxi}), we infer the relations
	\begin{equation}\label{GOma}
		\begin{array}{l}
			\smallskip
			\displaystyle \Omega_m(a)=\dfrac{1}{1+c_1 a^{3\left( \omega_m-						\omega_r \right)}+c_2a^{3\left( 1+\omega_m \right)}
			+c_3a^{ 1+3\omega_m }}, \\
			
			\smallskip
			\displaystyle \Omega_r(a)=\dfrac{c_1}{a^{3\left( 						\omega_r -\omega_m \right)}+c_1+c_2a^{3\left( 1+\omega_r \right)}
			+c_3a^{ 1+3\omega_r }},\\
			
			\smallskip
			\displaystyle \Omega_{\Lambda}(a)=\dfrac{c_2}{a^{-3\left( 1+						\omega_m \right)}+c_1a^{-3\left( 1+\omega_r \right)}+c_2
			+c_3a^{-2}},\\		
		\end{array}			
	\end{equation}	
and 
	\begin{equation}\label{shortG}
		\begin{array}{l}
			\smallskip \displaystyle \Omega_r(a)=c_1 a^{3\left( \omega_m-\omega_r 				\right)}\Omega_m(a),\\
			
			\smallskip \displaystyle \Omega_{\Lambda}(a)=c_2 a^{3 \left( 1+\omega_m 				\right)}\Omega_m(a),\\
			
			\smallskip \displaystyle \Omega_k(a)=1-\Omega_m(a) \left( 							1+c_1a^{3\left( \omega_m-\omega_r \right)}+c_2 a^{3 \left( 1+\omega_m 				\right)} \right).									
		\end{array}		
	\end{equation}

Values of the constants $c_1$, $c_2$ and $c_3$ we derive using the condition (\ref{sum}) calculated in the present time moment $t=t_0$, $a(t_0)=1$.	

For currently measured mean values of the density parameters $\Omega_i$, we use notation $\Omega_{i0}$. According to \cite{pdg}, those values are
	\begin{equation}\label{presentDG}
		\displaystyle \Omega_{m0} = 0.315, \quad \Omega_{r0} = 0.0000538, \quad 					\Omega_{k0} = 0.0007, \quad \Omega_{\Lambda0} = 0.685,
	\end{equation}	
where $\Omega_{m0}$ is the sum of mean values of baryon density and cold dark matter density of the universe, while for $\Omega_{r0}$ we take mean value of CMB radiation density of the universe. We note that some values in (\ref{presentDG}) are calculated independently, so the identity (\ref{sum}) is not conserved. However, that identity must hold in $\Lambda$CDM model, therefore here we use normalized values of $\Omega_{i0}$, denoted with $\Omega_{i0n}$, as
	\begin{equation}\label{normalG}
		\displaystyle \Omega_{i0n}=\dfrac{\Omega_{i0}}{\Omega_{m0}+\Omega_{r0}+				\Omega_{k0}+\Omega_{\Lambda0}},
	\end{equation}
where $\Omega_{i0}$ are values from (\ref{presentDG}).

Observing the system (\ref{GOma}) in the present time moment $t=t_0$,  $a(t_0)=1$, as well as substituting normalized present mean values of density parameters, we infer the following  
	\begin{equation}\label{solutionCG}
		c_1 \approx 0.000170794, \quad c_2 \approx 2.1746, \quad c_3 \approx 					0.00222148.
	\end{equation}

Replacing values of $c_1$, $c_2$ and $c_3$ from (\ref{solutionCG}) into (\ref{GOma}), we obtain an approximate solution of the system (\ref{systemG}).

We note that relations (\ref{GOma}) and (\ref{shortG}) represent new algebraic dependencies between density parameters, the scale factor and the equation of state parameters for matter and radiation. Also, we stress that every choice of equation of state parameters $\omega_m, \omega_r \in \left[ 0,1 \right] $, $\omega_m \neq \omega_r$, determine new cosmological model with cosmological constant $\Lambda$. 

Usually, in cosmology is supposed that $\omega_m=0$ and $\omega_r=1/3$. In accordance with that, we substitute those values of the equation of state parameters for matter and radiation in system (\ref{system}) in order to obtain
	\begin{equation}\label{systemN}
		\begin{array}{l}
			\smallskip
			\displaystyle {\Omega_m}'=\Omega_m \left(-1+\Omega_m +2\Omega_r 						-2\Omega_{\Lambda}\right),\\	
			
			\smallskip
			\displaystyle {\Omega_r}'=\Omega_r \left(-2+\Omega_m +2\Omega_r 						-2\Omega_{\Lambda}\right),\\
			
			\smallskip
			\displaystyle {\Omega_{\Lambda}}'=\Omega_{\Lambda} \left(2+\Omega_m 					+2\Omega_r-2\Omega_{\Lambda}\right),\\
			
			\smallskip
			\displaystyle \Omega_m+\Omega_r+\Omega_{\Lambda}+\Omega_k=1,\\
			
			\smallskip
			{\Omega_i}'=\dfrac{\dot{\Omega_i}}{H}.				
		\end{array}		
	\end{equation}

Substituting $\omega_m=0$ and $\omega_r=1/3$ in the equations (\ref{GOma}) and (\ref{shortG}), we obtain solution of the system (\ref{systemN})
 	\begin{equation}\label{Omaa}
		\begin{array}{l}
			\smallskip
			\displaystyle \Omega_m(a)=\dfrac{1}{1+c_1 a^{-1}+c_2a^3+c_3a}, \\
			
			\smallskip
			\displaystyle \Omega_r(a)=\dfrac{c_1}{a+c_1+c_2a^4+c_3a^2},\\
			
			\smallskip
			\displaystyle \Omega_{\Lambda}(a)=\dfrac{c_2}{a^{-3}+c_1a^{-4}+c_2
			+c_3a^{-2}},\\		
		\end{array}			
	\end{equation}
as well as	
	\begin{equation}\label{shorta}
		\begin{array}{l}
			\smallskip \displaystyle \Omega_r(a)=c_1 a^{-1}\Omega_m(a),\\
			
			\smallskip \displaystyle \Omega_{\Lambda}(a)=c_2 a^3 \Omega_m(a),\\
			
			\smallskip \displaystyle \Omega_k(a)=1-\Omega_m(a) \left( 1+c_1a^{-1}					+c_2 a^3 \right).									
		\end{array}		
	\end{equation}

 Furthermore, following the same treatment as in the general case, we obtain an approximate solution of the system (\ref{systemN}), with $c_1$, $c_2$ and $c_3$ taking values from (\ref{solutionCG}).

We note that a form of algebraic dependencies (\ref{Omaa}) and (\ref{shorta}) is obtained in \cite{dz}, with approach that is independent with the one presented here. In that way, an algebraic verification of this result is inferred in \cite{dz}.

\subsection{Two dimensional dynamical system}
\label{sec:2}

Here we consider that only one of four $\Omega_i$ is equal to zero. Therefore, we discuss four different cases. 

\subsubsection{Case $\Omega_m=0$, $\Omega_r>0$, $\Omega_{\Lambda}>0$ and $\Omega_k>0$}
\label{sec:RLK}

This case considers open universe, with both radiation and cosmological constant. Deficiency of matter fluid directly implies to the early universe's phase, since a term with properties of dark energy may caused rapid expansion of the early universe \cite{nied}, \cite{poulin}. Nevertheless, that theory is not consentient with the standard cosmological model. In spite of that, we analyze this case in order to obtain all solutions of the system (\ref{system}).

In this situation, the system (\ref{system}) has the following form

\begin{equation}\label{21dim}
		\begin{array}{l}
			\smallskip
			\displaystyle {\Omega_r}'=\Omega_r \left(-\left( 1+3\omega_r \right)  +\Omega_r \left(1+3\omega_r \right)-2\Omega_{\Lambda}											 \right),\\
			
			\smallskip
			\displaystyle {\Omega_{\Lambda}}'=\Omega_{\Lambda} \left(2+\Omega_r 													 \left(1+3\omega_r 																	 \right)-2\Omega_{\Lambda}\right),\\			
		\end{array}
	\end{equation}
with $\Omega_r+\Omega_{\Lambda}+\Omega_k=1$ and ${\Omega_i}'=\dfrac{\dot{\Omega_i}}{H}$. In the same manner as in Subsect.~\ref{sec:3}, we infer the solution of the system (\ref{21dim})
	\begin{equation}\label{sol21}
		\begin{array}{l}
			\smallskip
			\displaystyle \Omega_r(a)=\dfrac{1}{1+c_1 a^{3\left( 1+						\omega_r \right)}+c_2a^{ 1+3\omega_r }}, \\
			
			\smallskip
			\displaystyle \Omega_{\Lambda}(a)=\dfrac{c_1}{a^{-3\left( 1+						\omega_r \right)}+c_1+c_2a^{-2}},\\		
		\end{array}			
	\end{equation}	
where $c_1$ and $c_2$ are strictly positive constants. From (\ref{sol21}) and $\Omega_k=1-\Omega_r-\Omega_{\Lambda}$ follows 
	\begin{equation}\label{sol21alg}
		\begin{array}{l}
			\smallskip 
			\displaystyle \Omega_{\Lambda}(a)=c_1 a^{3 \left( 1+\omega_r 				\right)}\Omega_r(a),\\
			
			\smallskip 
			\displaystyle \Omega_k(a)=1-\Omega_r(a) \left( 							1+c_1 a^{3 \left( 1+\omega_r \right)}\right).						
		\end{array}		
	\end{equation}

\subsubsection{Case $\Omega_m>0$, $\Omega_r=0$, $\Omega_{\Lambda}>0$ and $\Omega_k>0$}
\label{sec:MLK}

According to a lack of radiation, this is an example of an open universe, around time of density equality of matter and dark energy, i.e. around $10^9$ yrs.  

Here, from the system (\ref{system}) we obtain

\begin{equation}\label{22dim}
		\begin{array}{l}
			\smallskip
			\displaystyle {\Omega_m}'=\Omega_m \left(-\left( 1+3\omega_m \right)  +\Omega_m \left(1+3\omega_m \right)-2\Omega_{\Lambda}											 \right),\\
			
			\smallskip
			\displaystyle {\Omega_{\Lambda}}'=\Omega_{\Lambda} \left(2+\Omega_m 													 \left(1+3\omega_m 																	 \right)-2\Omega_{\Lambda}\right),\\			
		\end{array}
	\end{equation}
with $\Omega_m+\Omega_{\Lambda}+\Omega_k=1$ and ${\Omega_i}'=\dfrac{\dot{\Omega_i}}{H}$. Following the same treatment as in Subsect.~\ref{sec:3}, we infer the solution of the system (\ref{22dim})
	\begin{equation}\label{sol22}
		\begin{array}{l}
			\smallskip
			\displaystyle \Omega_m(a)=\dfrac{1}{1+c_1 a^{3\left( 1+						\omega_m \right)}+c_2a^{ 1+3\omega_m }}, \\
			
			\smallskip
			\displaystyle \Omega_{\Lambda}(a)=\dfrac{c_1}{a^{-3\left( 1+						\omega_m \right)}+c_1+c_2a^{-2}},\\		
		\end{array}			
	\end{equation}	
where $c_1$ and $c_2$ are strictly positive constants. From (\ref{sol22}) and $\Omega_k=1-\Omega_m-\Omega_{\Lambda}$ follows 
	\begin{equation}\label{sol22alg}
		\begin{array}{l}
			\smallskip 
			\displaystyle \Omega_{\Lambda}(a)=c_1 a^{3 \left( 1+\omega_m 				\right)}\Omega_m(a),\\
			
			\smallskip 
			\displaystyle \Omega_k(a)=1-\Omega_m(a) \left( 							1+c_1 a^{3 \left( 1+\omega_m \right)}\right).						
		\end{array}		
	\end{equation}

Note that, if we replace index $m$ with $r$ in the relations (\ref{sol22}) and (\ref{sol22alg}), we obtain relations (\ref{sol21}) and (\ref{sol21alg}). That is because the systems (\ref{21dim}) and (\ref{22dim}) are symmetric with respect to indices $m$ and $r$. Moreover, the same symmetry holds in the system (\ref{system}).

\subsubsection{Case $\Omega_m>0$, $\Omega_r>0$, $\Omega_{\Lambda}=0$ and $\Omega_k>0$}
\label{sec:MRK}

Here we have an open universe around radiation and matter transition phase, since at that time dark energy could be neglected.

We consider the system

\begin{equation}\label{23dim}
		\begin{array}{l}
			\smallskip
			\displaystyle {\Omega_m}'=\Omega_m \left(-\left( 1+3\omega_m \right)  +\Omega_m \left(1+3\omega_m \right)+\Omega_r \left(1+3\omega_r \right)										 \right),\\
			
			\smallskip
			\displaystyle {\Omega_r}'=\Omega_r \left(-\left( 1+3\omega_r \right)  +\Omega_m \left(1+3\omega_m \right)+\Omega_r \left(1+3\omega_r \right)\right),\\
					
		\end{array}
	\end{equation}
with $\Omega_m+\Omega_r+\Omega_k=1$ and ${\Omega_i}'=\dfrac{\dot{\Omega_i}}{H}$. In the same manner as in Subsect.~\ref{sec:3}, we obtain the solution of the system (\ref{23dim})
	\begin{equation}\label{sol23}
		\begin{array}{l}
			\smallskip
			\displaystyle \Omega_m(a)=\dfrac{1}{1+c_1 a^{3\left(\omega_m-						\omega_r \right)}+c_2a^{ 1+3\omega_m }}, \\
			
			\smallskip
			\displaystyle \Omega_r(a)=\dfrac{c_1}{a^{3\left(\omega_r-						\omega_m \right)}+c_1+c_2a^{ 1+3\omega_r }}, \\	
		\end{array}			
	\end{equation}	
where $c_1$ and $c_2$ are strictly positive constants. From (\ref{sol23}) and $\Omega_k=1-\Omega_m-\Omega_r$ follows 
	\begin{equation}\label{sol23alg}
		\begin{array}{l}
			\smallskip 
			\displaystyle \Omega_r(a)=c_1 a^{3 \left(\omega_m-\omega_r 				\right)}\Omega_m(a),\\
			
			\smallskip \displaystyle \Omega_k(a)=1-\Omega_m(a) \left( 							1+c_1 a^{3 \left( \omega_m-\omega_r \right)}\right).						
		\end{array}		
	\end{equation}

\subsubsection{Case $\Omega_m>0$, $\Omega_r>0$, $\Omega_{\Lambda}>0$ and $\Omega_k=0$}
\label{sec:MRL}

We have a flat universe in the matter dominated stage, more precisely around $5 \cdot 10^8$ yrs (radiation and dark energy transition phase).

Substituting  $\Omega_{\Lambda}=1-\Omega_m-\Omega_r$ in (\ref{system}), we infer

\begin{equation}\label{24dim}
		\begin{array}{l}
			\smallskip
			\displaystyle {\Omega_m}'=\Omega_m \left(-3\left( 1+\omega_m \right)  +3\left(1+\omega_m \right)\Omega_m +3\left(1+\omega_r \right)\Omega_r 										 \right),\\
			
			\smallskip
			\displaystyle {\Omega_r}'=\Omega_r \left(-3\left( 1+\omega_r \right)  +3\left(1+\omega_m \right)\Omega_m +3\left(1+\omega_r \right)\Omega_r 										 \right),\\
					
		\end{array}
	\end{equation}
with ${\Omega_i}'=\dfrac{\dot{\Omega_i}}{H}$. Following the procedure presented in Subsect.~\ref{sec:3}, we obtain the solution of the system (\ref{24dim})
	\begin{equation}\label{sol24}
		\begin{array}{l}
			\smallskip
			\displaystyle \Omega_m(a)=\dfrac{1}{1+c_1 a^{3\left(\omega_m-						\omega_r \right)}+c_2a^{3\left( 1+3\omega_m \right)}}, \\
			
			\smallskip
			\displaystyle \Omega_r(a)=\dfrac{c_1}{a^{3\left(\omega_r-						\omega_m \right)}+c_1+c_2a^{ 3\left(1+3\omega_r \right)}}, \\	
		\end{array}			
	\end{equation}	
where $c_1$ and $c_2$ are strictly positive constants. From (\ref{sol24}) and $\Omega_{\Lambda}=1-\Omega_m-\Omega_r$ follows 
	\begin{equation}\label{sol24alg}
		\begin{array}{l}
			\smallskip 
			\displaystyle \Omega_r(a)=c_1 a^{3 \left(\omega_m-\omega_r 				\right)}\Omega_m(a),\\
			
			\smallskip 
			\displaystyle \Omega_{\Lambda}(a)=1-\Omega_m(a) \left( 							1+c_1 a^{3 \left( \omega_m-\omega_r \right)}\right).						
		\end{array}		
	\end{equation}

\subsection{One dimensional dynamical system}
\label{sec:1}

Finally, we observe the system (\ref{system}) when exactly two of four $\Omega_i$ are equal to zero. That lead to six different cases. In each case, the dynamical system consists of only one differential equation.

\subsubsection{Case $\Omega_m=0$, $\Omega_r=0$, $\Omega_{\Lambda}>0$ and $\Omega_k>0$}
\label{sec:LK}

Taking into account a lack of matter and radiation, as well as dark energy domination, this is a future stage of an open universe.
 
Substituting $\Omega_m=0$ and $\Omega_r=0$ into the system (\ref{system}), we obtain
	\begin{equation}\label{11dim}
		\displaystyle {\Omega_{\Lambda}}'=\Omega_{\Lambda}(2-2\Omega_{\Lambda}),
	\end{equation}
with $\Omega_{\Lambda}+\Omega_k=1$ and ${\Omega_{\Lambda}}'=\dfrac{\dot{\Omega_{\Lambda}}}{H}$. The solution of the differential equation (\ref{11dim}) is
	\begin{equation}\label{sol11}
		\displaystyle \Omega_{\Lambda}(a)=\dfrac{a^2}{a^2+c_1},	
	\end{equation}
where $c_1$ is strictly positive constant. From $\Omega_k=1-\Omega_{\Lambda}$ and (\ref{sol11}), follows \newline $\Omega_k(a)=\dfrac{c_1}{a^2+c_1}$.
	
\subsubsection{Case $\Omega_m=0$, $\Omega_r>0$, $\Omega_{\Lambda}=0$ and $\Omega_k>0$}
\label{sec:RK}

In this case, we analyze an open universe with radiation as a dominant fluid, i.e. an early phase of an open universe.

Replacing $\Omega_m=0$ and $\Omega_{\Lambda}=0$ into the system (\ref{system}), we infer
	\begin{equation}\label{12dim}
		\displaystyle {\Omega_r}'=\Omega_r \left(-\left( 1+3\omega_r \right)+\Omega_r \left( 1+3\omega_r \right) \right),
	\end{equation}
with $\Omega_r+\Omega_k=1$ and ${\Omega_r}'=\dfrac{\dot{\Omega_r}}{H}$. Solving (\ref{12dim}) and using $\Omega_k=1-\Omega_r$, we derive
	\begin{equation}\label{sol12}
		\begin{array}{l}
			\smallskip
			\displaystyle \Omega_r(a)=\dfrac{1}{1+c_1 a^{1+3\omega_r}},\\
			
			\smallskip
			\displaystyle \Omega_k(a)=1-\dfrac{1}{1+c_1 a^{1+3\omega_r}},\\
			
		\end{array}	
	\end{equation}
where $c_1$ is strictly positive constant.

\subsubsection{Case $\Omega_m=0$, $\Omega_r>0$, $\Omega_{\Lambda}>0$ and $\Omega_k=0$}
\label{sec:RL}

Here we have flat universe with radiation and dark energy. For the same reason as in 3.2.1, we analyze this case only for the sake of completeness of this manuscript, despite the absence of physical interpretation in the context of the $\Lambda$CDM model.

 Here we observe 
	\begin{equation}\label{13dim}
		\displaystyle {\Omega_r}'=\Omega_r \left(-3\left( 1+\omega_r \right)+3\left( 1+\omega_r \right)\Omega_r  \right),
	\end{equation}
with $\Omega_r+\Omega_{\Lambda}=1$ and ${\Omega_r}'=\dfrac{\dot{\Omega_r}}{H}$. Solving (\ref{13dim}) and replacing its solution into $\Omega_{\Lambda}=1-\Omega_r$, we obtain
	\begin{equation}\label{sol13}
		\begin{array}{l}
			\smallskip
			\displaystyle \Omega_r(a)=\dfrac{1}{1+c_1 a^{3\left(1+\omega_r\right)}},\\
			
			\smallskip
			\displaystyle \Omega_{\Lambda}(a)=1-\dfrac{1}{1+c_1 a^{3\left(1+\omega_r\right)}},\\
			
		\end{array}	
	\end{equation}
where $c_1$ is strictly positive constant.

\subsubsection{Case $\Omega_m>0$, $\Omega_r=0$, $\Omega_{\Lambda}=0$ and $\Omega_k>0$}
\label{sec:MK}

The universe in the description is open and in matter dominated phase.

Since the system (\ref{system}) is symmetric with respect to indices $m$ and $r$, the solution of the differential equation
\begin{equation}\label{14dim}
		\displaystyle {\Omega_m}'=\Omega_m \left(-\left( 1+3\omega_m \right)+\Omega_m \left( 1+3\omega_m \right) \right),
	\end{equation}
with $\Omega_m+\Omega_k=1$ and ${\Omega_m}'=\dfrac{\dot{\Omega_m}}{H}$,
is the same as the solution of the differential equation (\ref{12dim}), after replacing index $r$ with $m$, i.e.
	\begin{equation}\label{sol14}
		\begin{array}{l}
			\smallskip
			\displaystyle \Omega_m(a)=\dfrac{1}{1+c_1 a^{1+3\omega_m}},\\
			
			\smallskip
			\displaystyle \Omega_k(a)=1-\dfrac{1}{1+c_1 a^{1+3\omega_m}},\\
			
		\end{array}	
	\end{equation}
where $c_1$ is strictly positive constant.

\subsubsection{Case $\Omega_m>0$, $\Omega_r=0$, $\Omega_{\Lambda}>0$ and $\Omega_k=0$}
\label{sec:ML}

Since the presence of both matter and cosmological constant in a flat universe, as well as the absence of radiation, this suits the present stage of our universe.

Again, according to the already mentioned symmetry between the indices $m$ and $r$, the solution of the differential equation 
	\begin{equation}\label{15dim}
		\displaystyle {\Omega_m}'=\Omega_m \left(-3\left( 1+\omega_m \right)+3\left( 1+\omega_m \right)\Omega_m  \right),
	\end{equation}
with $\Omega_m+\Omega_{\Lambda}=1$ and ${\Omega_m}'=\dfrac{\dot{\Omega_m}}{H}$, is the same as the solution of the differential equation (\ref{13dim}), after replacing index $r$ with $m$, i.e.
	\begin{equation}\label{sol15}
		\begin{array}{l}
			\smallskip
			\displaystyle \Omega_m(a)=\dfrac{1}{1+c_1 a^{3\left(1+\omega_m\right)}},\\
			
			\smallskip
			\displaystyle \Omega_{\Lambda}(a)=1-\dfrac{1}{1+c_1 a^{3\left(1+\omega_m\right)}},\\
			
		\end{array}	
	\end{equation}
where $c_1$ is strictly positive constant.

\subsubsection{Case $\Omega_m>0$, $\Omega_r>0$, $\Omega_{\Lambda}=0$ and $\Omega_k=0$}
\label{sec:MR}

In this case, we deal with a flat universe around a time period of transition between radiation and matter.

Here, we solve
	\begin{equation}\label{16dim}
		\displaystyle {\Omega_m}'=\Omega_m \left(-3\left( \omega_m - \omega_r\right)		+3\left(\omega_m-\omega_r \right)\Omega_m  \right),
	\end{equation}
with $\Omega_m+\Omega_r=1$ and ${\Omega_m}'=\dfrac{\dot{\Omega_m}}{H}$. Solving previous differential equation and substituting its solution into $\Omega_r=1-\Omega_m$, we infer
	\begin{equation}\label{sol16}
		\begin{array}{l}
			\smallskip
			\displaystyle \Omega_m(a)=\dfrac{a^{3\omega_r}}{a^{3\omega_r}+c_1 a^{3\omega_m}},\\
			
			\smallskip
			\displaystyle \Omega_r (a)=1-\dfrac{a^{3\omega_r}}{a^{3\omega_r}+c_1 a^{3\omega_m}},\\
			
		\end{array}	
	\end{equation}
where $c_1$ is strictly positive constant.

In various literature (for example see \cite{ds}, \cite{jungle}, \cite{garcia}, \cite{uzan}) the system (\ref{systemN}) or some similar systems are analyzed by applying theory of dynamical systems. We present that method here in more details in order to analyze the system (\ref{system}), which is more general. That analysis enable us to clearly understand the evolution of the universe from new perspective. We will see how density parameters ruled the expansion of the universe from its beginning and how their behaviour will have direct impact on the universe's future. 

\section{Evolution of the dynamical system}
\label{sec:evolution}

\begin{figure*}
\centering
\includegraphics[width=0.75\textwidth]{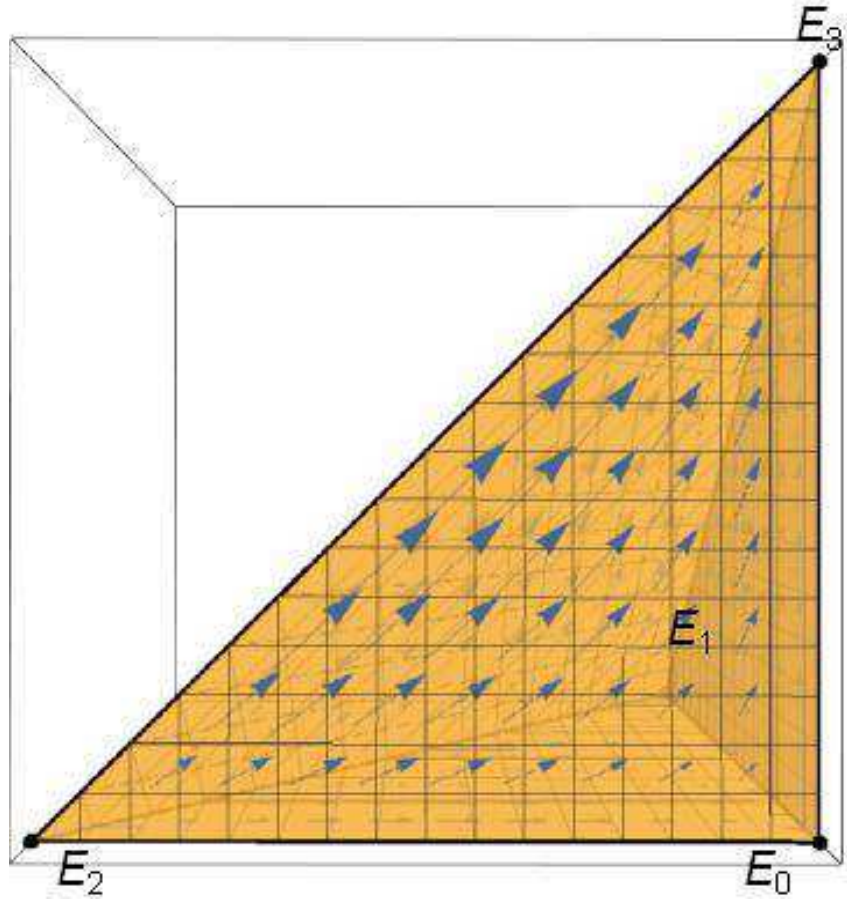}
\caption{Phase space portrait of the dynamical system (\ref{system}) with the values $\omega_m=0$ and $\omega_r=1/3$.}
\label{fig:Graph1}
\end{figure*}

According to $\Omega_i \in \left[ 0,1 \right]$ and (\ref{sum}), the phase space for the system (\ref{system}) is tetrahedron whose vertices are points $E_0 =\left( 0,0,0\right)$, $E_1 =\left( 1,0,0\right)$, $E_2 =\left( 0,1,0\right)$ and $E_3 =\left( 0,0,1\right)$, which lie on the axes $\Omega_m$, $\Omega_r$ and $\Omega_{\Lambda}$, (Figure~\ref{fig:Graph1}, see  \cite{figures}).

Since $\omega_m, \omega_r \in \left[ 0,1 \right]$ and $\omega_m \neq \omega_r$, the only equilibriums of the system (\ref{system}) are exactly the vertices $E_0 =\left( 0,0,0\right)$, $E_1 =\left( 1,0,0\right)$, $E_2 =\left( 0,1,0\right)$ and $E_3 =\left( 0,0,1\right)$. Now we apply linear stability analysis near the equilibrium points in order to obtain the behaviour of the universe. 

\noindent{\bf{The equilibrium \boldmath{$E_0$}.}} The eigenvalues of the Jacobian at the critical point $E_0$ are $\lambda_1=-\left( 1+3\omega_m \right)<0$, $\lambda_2=-\left( 1+3\omega_r \right)<0$ and $\lambda_3=2$, wherefrom we conclude that $E_0$ is a saddle point.
Considering that $E_0$ is the origin, i.e. that $\Omega_m=0$, $\Omega_r=0$ and $\Omega_{\Lambda}=0$, we infer $\Omega_k=1$, wherefrom $k<0$. Hence, the universe near the equilibrium $E_0$ is empty and open universe without $\Lambda$, known as the Milne universe. According to the discussion above, we conclude that this is an unstable phase of the universe's evolution.
\medskip

\noindent{\bf{The equilibrium \boldmath{$E_1$}.}} When it comes to the critical point $E_1$, the eigenvalues of the Jacobian are $\lambda_1= 1+3\omega_m>0$, $\lambda_2=3\left(\omega_m-\omega_r \right) \neq 0$ and $\lambda_3=3\left( 1+\omega_m \right)>0$. We have two cases, depending on the sign of $\left(\omega_m-\omega_r\right)$. If $\omega_m>\omega_r$, then $\lambda_2>0$, which implies that $E_1$ is an unstable point (repeller). Otherwise, $E_1$ is a saddle point. According to the coordinates of $E_1$, i.e. that $\Omega_m=1$, $\Omega_r=0$ and $\Omega_{\Lambda}=0$, we infer $\Omega_k=0$, wherefrom $k=0$. Therefore, the universe near the equilibrium $E_1$ is flat universe and matter dominated, without $\Lambda$, known as the Einstein-de Sitter universe. We note that, regardless of the sign of $\left(\omega_m-\omega_r\right)$, this phase of the universe's evolution is unstable. 
\medskip

\noindent{\bf{The equilibrium \boldmath{$E_2$}.}} The eigenvalues of the Jacobian at the equilibrium $E_2$ are symmetric to the eigenvalues at $E_1$, with respect to the indices $m$ and $r$, i.e $\lambda_1= 1+3\omega_r>0$, $\lambda_2=3\left(\omega_r-\omega_m \right) \neq 0$ and $\lambda_3=3\left( 1+\omega_r \right)>0$. Again, the nature of the equilibrium depends on the sign of $\left(\omega_m-\omega_r\right)$. Here, if $\omega_m>\omega_r$, then $\lambda_2<0$, which indicates that $E_2$ is a saddle point. Otherwise, $E_2$ is an unstable point (repeller).
Since in this case is $\Omega_m=0$, $\Omega_r=1$ and $\Omega_{\Lambda}=0$, we infer $\Omega_k=0$, as well as $k=0$. To conclude, the universe near the critical point $E_2$ is flat universe and radiation dominated, without cosmological constant. As for the previous case, the universe near this critical point is unstable.
\medskip

\noindent{\bf{The equilibrium \boldmath{$E_3$}.}} Finally, the eigenvalues of the Jacobian at the critical point $E_3$ are $\lambda_1=-3\left( 1+\omega_m \right)<0$, $\lambda_2=-3\left( 1+\omega_r \right)<0$ and $\lambda_3=-2$, wherefrom $E_3$ is a stable point (attractor).
Here we have $\Omega_m=0$, $\Omega_r=0$, $\Omega_{\Lambda}=1$ and again $k=0$. Hence, the universe near the equilibrium $E_3$ is flat and $\Lambda$-dominated, also known as the de-Sitter universe. Thus, this is a stable phase in the evolution of the universe.
\medskip

To summarize, $E_0$ is a saddle point, $E_3$ is an attractor, while the nature of $E_1$ and $E_2$ depends on the sign of $\left(\omega_m-\omega_r\right)$. Without loss of generality, we analyze the case $\omega_m<\omega_r$. Thus, $E_1$ is a saddle point, whereas $E_2$ is a repeller.

Solutions of the system (\ref{system}) are the orbits that start at $E_2$ as $\xi \rightarrow -\infty$, i.e. $a(t)\rightarrow 0_{+} $, and end in $E_3$ as $\xi \rightarrow +\infty$, i.e. $a(t)\rightarrow +\infty$. It means that the early universe was flat and radiation dominated, while the future of the flat universe is characterised by $\Lambda$-domination, which is consistent with the observations. The exceptions are the orbits on the boundaries and in the interior of the triangles $\Delta E_0E_1E_2$ and $\Delta E_0E_1E_3$, that lie in the ($\Omega_m$,$\Omega_r$)-plane and ($\Omega_m$,$\Omega_{\Lambda}$)-plane, respectively. Those exceptions are expected, since they correspond to the trajectories in a universe without cosmological constant, or to the trajectories in a universe without radiation, respectively. In the following, we analyze these situations in more details. 

\subsection{A universe without cosmological constant $\Lambda$} 
\label{sec:WL}

Replacing $\Omega_{\Lambda}=0$ into (\ref{system}), we obtain
\begin{equation}\label{e23dim}
		\begin{array}{l}
			\smallskip
			\displaystyle {\Omega_m}'=\Omega_m \left(-\left( 1+3\omega_m \right)  +\Omega_m \left(1+3\omega_m \right)+\Omega_r \left(1+3\omega_r \right)										 \right),\\
			
			\smallskip
			\displaystyle {\Omega_r}'=\Omega_r \left(-\left( 1+3\omega_r \right)  +\Omega_m \left(1+3\omega_m \right)+\Omega_r \left(1+3\omega_r \right)\right),\\
					
		\end{array}
	\end{equation}
with $\Omega_i \in \left[ 0,1 \right]$, $\Omega_m+\Omega_r+\Omega_k=1$ and ${\Omega_i}'=\dfrac{\dot{\Omega_i}}{H}$, which is the same system as (\ref{23dim}). The solutions of the system (\ref{e23dim}) are derived in Subsubsects.~\ref{sec:MRK}, \ref{sec:RK}, \ref{sec:MK} and \ref{sec:MR}, depending on whether all $\Omega_i>0$, or otherwise.

\begin{figure*}
\centering
\includegraphics[width=0.75\textwidth]{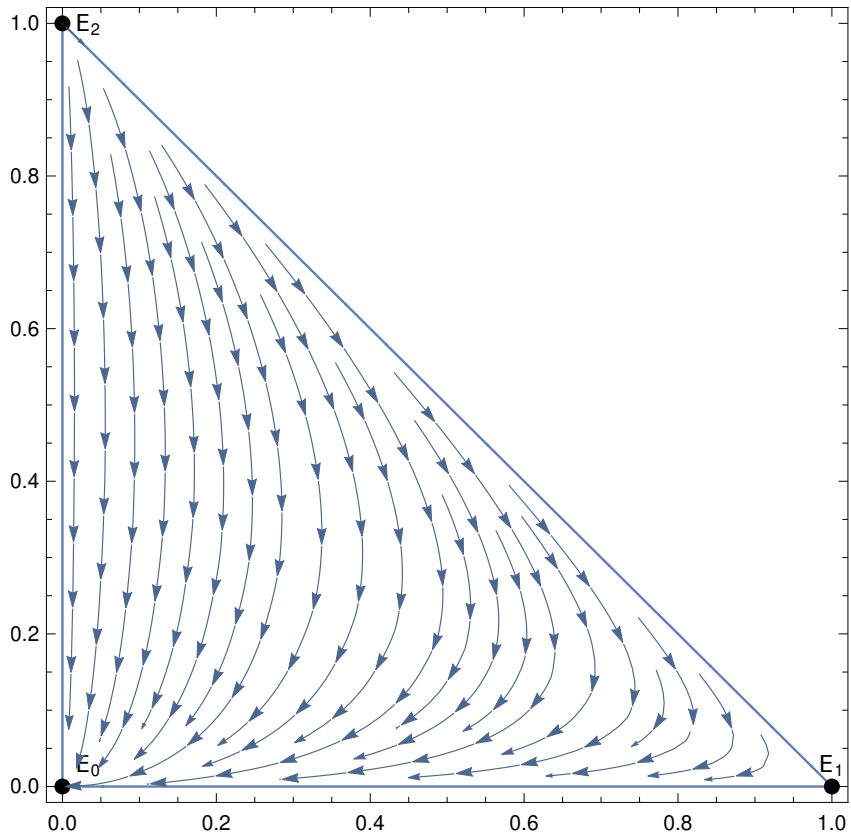}
\caption{Phase space portrait of the dynamical system (\ref{e23dim}) with the values $\omega_m=0$ and $\omega_r=1/3$.}
\label{fig:Graph2}
\end{figure*}

The phase space for the system (\ref{e23dim}) is the triangle in the $\left(\Omega_m,\Omega_r \right)$-plane whose vertices are points with coordinates $\left(0,0\right)$, $\left(1,0\right)$ and $\left(0,1\right)$, which are exactly the coordinates of the points $E_0$, $E_1$ and $E_2$ in the plane $\Omega_{\Lambda}=0$. Therefore, for the sake of simplicity we denote our phase space with $\Delta E_0E_1E_2$ (Figure~\ref{fig:Graph2}, see  \cite{figures}).

According to $\omega_m, \omega_r \in \left[ 0,1 \right]$ and $\omega_m < \omega_r$, the equilibriums of the system (\ref{e23dim}) are vertices $E_0 =\left( 0,0\right)$, $E_1 =\left( 1,0\right)$, and $E_2 =\left( 0,1\right)$. Utilizing linear stability analysis near the equilibrium points, we infer the following.

\noindent{\bf{The equilibrium \boldmath{$E_0$}.}} The eigenvalues of the Jacobian at $E_0$ are $\lambda_1=-\left( 1+3\omega_m \right)<0$ and $\lambda_2=-\left( 1+3\omega_r \right)<0$, wherefrom we conclude that $E_0$ is an attractor.
Similarly as in the three dimensional case, we conclude that the universe near the equilibrium $E_0$ is the Milne universe (empty and open universe without $\Lambda$). 
\medskip

\noindent{\bf{The equilibrium \boldmath{$E_1$}.}} In this case, the eigenvalues of the Jacobian are $\lambda_1= 1+3\omega_m>0$ and $\lambda_2=3\left(\omega_m-\omega_r \right) < 0$. Consequently, $E_1$ is a saddle point. Furthermore, a universe near critical point $E_1$ is Einstein-de Sitter universe (flat universe and matter dominated, without $\Lambda$). 
\medskip

\noindent{\bf{The equilibrium \boldmath{$E_2$}.}} Finally, $E_2$ is a repeller, given that the eigenvalues of the Jacobian in this case are $\lambda_1= 1+3\omega_r>0$ and $\lambda_2=3\left(\omega_r-\omega_m \right) > 0$. The universe near the critical point $E_2$ is flat universe and radiation dominated, without cosmological constant.

In conclusion, solutions of the system (\ref{e23dim}) are the orbits that start at $E_2$ as $\xi \rightarrow -\infty$ and end in $E_0$ as $\xi \rightarrow +\infty$. That is quite interesting, because it states that in the absence of $\Lambda$, nearly flat and radiation dominated universe will tend to evolve into the Milne universe, which is open and empty universe. The exceptions are the orbits that connect $E_2$ to $E_1$ and $E_1$ to $E_0$, i.e. the ones on the line $\Omega_r=1-\Omega_m$ and on the $\Omega_m$-axis. These orbits correlate either to a flat universe without $\Lambda$, or to a universe without radiation and $\Lambda$. 
In the first case, they describe the tendency of evolving flat and radiation dominated universe without cosmological constant into the Einstein-de Sitter universe, or evolving the Einstein-de Sitter universe into the Milne universe in the latter case.

\subsection{A universe without radiation}
\label{sec:WR}

Substituting $\Omega_r=0$ into (\ref{system}), we infer
\begin{equation}\label{e22dim}
		\begin{array}{l}
			\smallskip
			\displaystyle {\Omega_m}'=\Omega_m \left(-\left( 1+3\omega_m \right)  +\Omega_m \left(1+3\omega_m \right)-2\Omega_{\Lambda}											 \right),\\
			
			\smallskip
			\displaystyle {\Omega_{\Lambda}}'=\Omega_{\Lambda} \left(2+\Omega_m 													 \left(1+3\omega_m 																	 \right)-2\Omega_{\Lambda}\right),\\			
		\end{array}
	\end{equation}
with $\Omega_m+\Omega_{\Lambda}+\Omega_k=1$ and ${\Omega_i}'=\dfrac{\dot{\Omega_i}}{H}$, which is the same system as (\ref{22dim}). The solutions of the system (\ref{e22dim}) are obtained in Subsubsects.~\ref{sec:MLK}, \ref{sec:LK}, \ref{sec:MK} and \ref{sec:ML}, depending on whether all $\Omega_i>0$, or otherwise.

\begin{figure*}
\centering
\includegraphics[width=0.75\textwidth]{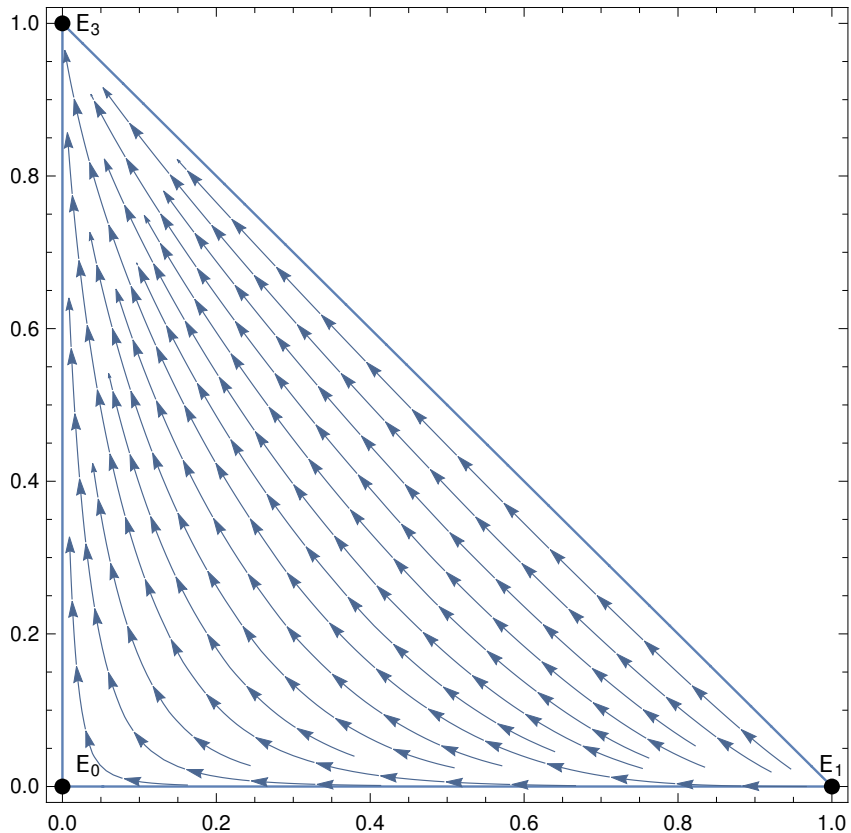}
\caption{Phase space portrait of the dynamical system (\ref{e22dim}) with the value $\omega_m=0$.}
\label{fig:Graph3}
\end{figure*}

The phase space for the system (\ref{e22dim}) is the triangle in the $\left(\Omega_m,\Omega_{\Lambda} \right)$-plane whose vertices are points with coordinates $\left(0,0\right)$, $\left(1,0\right)$ and $\left(0,1\right)$, which are exactly the coordinates of the points $E_0$, $E_1$ and $E_3$ in the plane $\Omega_r=0$. Thus, for the sake of simplicity we denote our phase space with $\Delta E_0E_1E_3$ (Figure~\ref{fig:Graph3}, see  \cite{figures}).

Since $\omega_m \in \left[ 0,1 \right]$, the equilibriums of the system (\ref{e22dim}) are the vertices $E_0 =\left( 0,0\right)$, $E_1 =\left( 1,0\right)$, and $E_3 =\left( 0,1\right)$. Using linear stability analysis near the equilibrium points, we infer the following.
\medskip

\noindent{\bf{The equilibrium \boldmath{$E_0$}.}} In this case, the eigenvalues of the Jacobian at $E_0$ are $\lambda_1=-\left( 1+3\omega_m \right)<0$ and $\lambda_2=2$. Consequently, $E_0$ is a saddle. In the same manner as in the three dimensional case, we conclude that the universe near the equilibrium $E_0$ is the Milne universe (empty and open universe without $\Lambda$). 
\medskip

\noindent{\bf{The equilibrium \boldmath{$E_1$}.}} The equilibrium $E_1$ is a repeller, given that the eigenvalues of the Jacobian in this case are $\lambda_1= 1+3\omega_m>0$ and $\lambda_2=3\left(1+\omega_m\right) > 0$. The universe near critical point $E_1$ is the Einstein-de Sitter universe (flat universe and matter dominated, without $\Lambda$). 
\medskip

\noindent{\bf{The equilibrium \boldmath{$E_3$}.}} The eigenvalues of the Jacobian are $\lambda_1=-2$ and \newline $\lambda_2= -3\left(1+\omega_m\right)<0$, wherefrom we conclude that $E_3$ is an attractor. Moreover, the universe near the critical point $E_3$ is the de-Sitter universe (flat universe and $\Lambda$ dominated).

Solutions of the system (\ref{e22dim}) are the orbits that start at $E_1$ as $\xi \rightarrow -\infty$ and end in $E_3$ as $\xi \rightarrow +\infty$. It implies that the Einstein-de Sitter universe  will tend to evolve to the de-Sitter universe. The exceptions are the orbits that connect $E_1$ to $E_0$ and $E_0$ to $E_3$, i.e. the ones on the $\Omega_m$-axis and on the $\Omega_{\Lambda}$-axis. These orbits correlate either to a universe without radiation and $\Lambda$, or to a universe without radiation and matter. 
In the first case, they describe the tendency of evolving the Einstein-de Sitter universe into the Milne universe (we noticed this in Subsect.~\ref{sec:WL}), or evolving the Milne universe into the de-Sitter universe in the latter case. 

\section{Conclusion}

In this paper we discussed the dynamics of the universe in the frame of the $\Lambda$CDM model. We proved the equivalence between the system of the Friedmann equations (\ref{sf}) and the dynamical system (\ref{system}), under some assumptions that are standard in cosmology. Furthermore, all solutions of the system (\ref{system}) are derived and they represent new parametrizations of the density parameters with respect to the scale factor. All solutions of the system (\ref{system}) are connected with appropriate stage of the universe's evolution. We demonstrated that linear stability analysis near the equilibrium points indicate to some interesting behaviour regarding to the dynamics of the universe. 

Our future work is directed to generalisation of the results in this paper for the universe filled with $n$ barotropic perfect fluids without mutual interaction. It will also be interesting to consider the universe with barotropic perfect coupled fluids. 
\medskip

\noindent{\bf{Note.}} All computations are checked using the Wolfram Mathematica package.
\medskip

\noindent{\bf{Acknowledgements.}} The authors would like to thank Jelena Kati\' c and Jovana Nikoli\'c for careful reading of the paper and the constructive and helpful remarks.

  \end{document}